\documentclass[namedreferences]{SolarPhysics}
\usepackage[optionalrh,solaenum]{spr-sola-addons} 
\usepackage{graphicx}        
\usepackage{color}           
\usepackage{url}             

\begin{document}

\begin{article}

\begin{opening}

\title{Simultaneous Observations of the Chromosphere with TRACE and SUMER\\ {\it Solar Physics}}

\author{Jay M. Pasachoff$^{1,2}$\sep
        Evan D. Tingle$^{3,4}$\sep
	Ingolf E. Dammasch$^{5}$\sep
	Alphonse C. Sterling$^{6,7}$}
\runningauthor{J. M. Pasachoff, E. D. Tingle, I. E. Dammasch, A.C. Sterling}
\runningtitle{Observations with TRACE and SUMER}

   \institute{$^{1}$ Williams College-Hopkins Observatory, Williamstown, MA 01267, USA email: \url{jay.m.pasachoff@williams.edu}\\ 
		$^{2}$  California Institute of Technology 150-21, Pasadena, CA 91125, USA\\
		$^{3}$ Wesleyan University-Van Vleck Observatory, Middletown, CT 06459, USA\\
		$^{4}$ Harvard-Smithsonian Center for Astrophysics, Cambridge, MA 02138\\
		$^{5}$ Royal Observatory of Belgium/Solar Influences Data Analysis Center (SIDC), 1180 Brussels, Belgium\\
		$^{6}$ NASA/MSFC, VP62/Space Science Office, Huntsville, AL 35805\\
		$^{7}$ Current address: JAXA/Institute of Space and Astronautical Science, \emph{Hinode} Group, 3-1-1 Yoshinodai, Sagamihara, Kanagawa 229-8510, Japan\\
}

\begin{abstract}
Using mainly the 1600 \AA\ continuum channel, and also the 1216 \AA\ Lyman-$\alpha$ channel (which includes some UV continuum and C {\sc iv} emission), aboard the TRACE satellite, we observed the complete lifetime of a transient, bright chromospheric loop. Simultaneous observations with the SUMER instrument aboard the SOHO spacecraft revealed interesting material velocities through the Doppler effect existing above the chromospheric loop imaged with TRACE, possibly corresponding to extended non-visible loops, or the base of an X-ray jet.
  
\end{abstract}
\end{opening}

\section{Introduction}
     \label{S-Introduction} 

During 24-27 May 2004, we performed a series of simultaneous chromospheric observations of the northwest and southwest solar limb with the SUMER (Solar Ultraviolet Measurements of Emitted Radiation) instrument aboard the SOHO (Solar and Heliospheric Observatory) spacecraft and with the TRACE (Transition Region and Coronal Explorer) satellite. This campaign was part of our more general recent program examining the solar limb (Pasachoff, Jacobson, and Sterling, 2009). Among the most prominent findings in our SUMER data was a transient feature in one of the spectral scans that showed strong concurrent red- and blue-shifted material velocities. Concurrent observations from TRACE revealed that the flows corresponded to a bright, transient chromospheric loop near the boundary of a polar coronal hole. Here we describe the observed properties of this feature, and speculate on its physical nature.

\section{Instrumentation} 
      \label{S-instrumentation}      
The SUMER instrument aboard the ESA/NASA SOHO spacecraft is a high-precision spectrometer with the ability to measure material velocities through the Doppler effect (Wilhelm \emph{et al.}, 1995).  For each day of the period mentioned above, the SUMER spectrometer took an approximately 90 min observation of both the northwest and southwest solar limbs in a sit-and-stare mode.  The orientation of the SUMER spectral slit was fixed in a north-south fashion, so an approximate 1 arcsec $\times$ 300 arcsec region of the sun was observed with the solar limb appearing around the solar $y$=$\pm$918 arcsec. SUMER observed three spectral windows surrounding three target wavelengths---Si {\sc ii}, C {\sc iv}, and Ne {\sc viii}---that correspond to different regions of the solar atmosphere.

\begin{table}[!h]
\begin{tabular}{cccc}

\hline
Ne {\sc viii} & 770.4 \AA & 630,000 K & Bottom of corona \\
\hline
C {\sc iv} & 1548.2 \AA & 100,000 K & Transition region \\
\hline
Si {\sc ii} & 1533.4 \AA & 14,000 K & Chromosphere \\
\hline
Photospheric cont.& Various wavelengths & 6,000 K & Photosphere \\
\hline
\end{tabular}
\caption{The wavelengths observed with SUMER with corresponding temperature and region of the solar atmosphere where the wavelength is predominately observed.}
\label{T-SUMER}
\end{table}

Each 1 arcsec $\times$ 1 arcsec SUMER spatial pixel in this study consists of 50 spectral pixels that correspond to a window of approximately 2 \AA, ideally containing a line profile with some adjacent continuum. Line and continuum pixels are identified for the whole study, and the continuum is estimated and subtracted from each profile. The line radiance is estimated by integrating all remaining line pixels, and the line position is estimated by the center of gravity of the line profile. The line shift is then estimated by the line position relative to the average position over the whole study.

NASA's TRACE spacecraft provides high-resolution images in the UV and EUV, combined with a quick cadence of approximately 40 s (Handy \emph{et al.}, 1999; Golub and Pasachoff, 2001, 2008). During the SUMER scans, simultaneous TRACE observations were made with alternating images taken in the 1600 \AA\ continuum filter and in the Lyman-$\alpha$ filter (1216 \AA), each with a cadence of approximately 40 s. (The Lyman-$\alpha$ channel includes some UV continuum and C {\sc iv} emission.) These images in quick succession allowed us to create movies to observe the evolution of the chromospheric feature throughout its lifetime. The transient chromospheric loop on which this project focused was only faintly visible in the Lyman-$\alpha$ images, so the 1600 \AA\ images were used primarily.

\section{Observations and Results} 
      \label{S-ObsResults}      

A special-purpose program was written to calculate the position of the solar limb, search the approximately 11 h of SUMER data in each wavelength looking for velocity spikes, and graphically display the information. As this project was originally designed to find and measure spicules, whenever the analysis software revealed a large velocity peak that occurred simultaneously in multiple adjacent $y$-coordinates corresponding to a coherent velocity spike throughout multiple heights of chromospheric material, that feature would be designated as a ``spicule candidate.'' Although not related to spicules, one particularly interesting measurement showed strong lineshifts in both Si {\sc ii} and C {\sc iv}.  After viewing the corresponding TRACE observations, it was discovered that these prominent velocity shifts were the product of a small, short-lived chromospheric loop.

\subsection{TRACE and EIT} 
  \label{S-TRACE-EIT}

Although we identified our target feature through its prominent signature in the SUMER data, we first discuss the observations in the TRACE data, since they provide a more general context for the observations.

  \begin{figure}   

\centering
\includegraphics[width=1.0\textwidth]{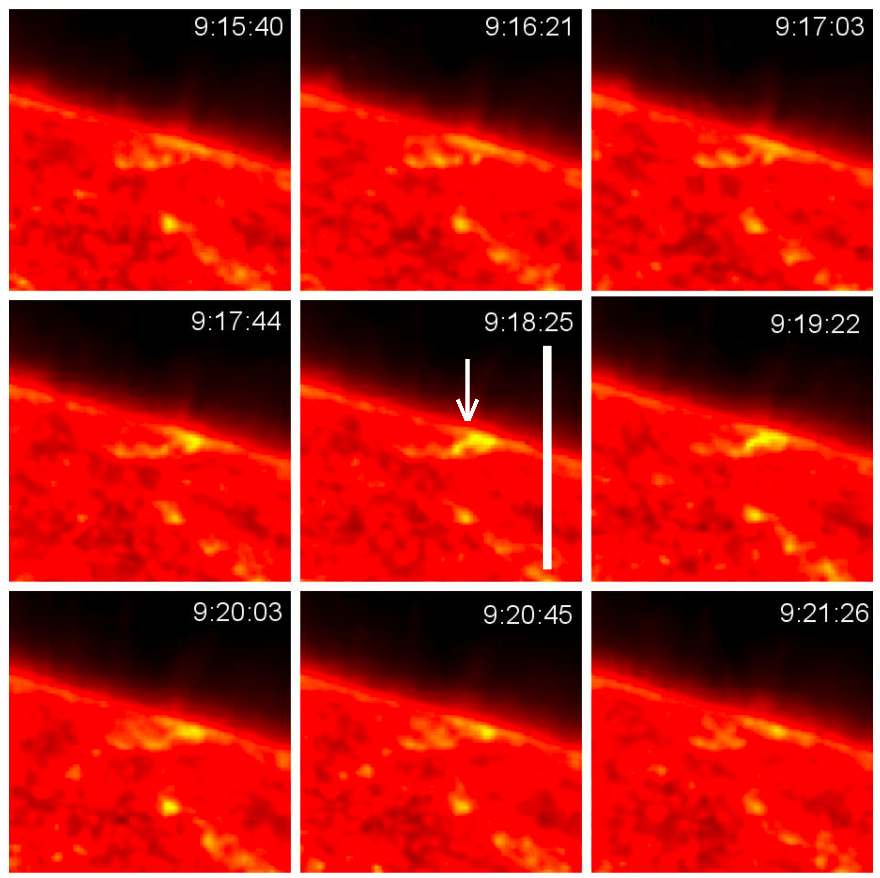}
\caption{This progression of TRACE images taken in the continuum wavelength of 1600 \AA\ on 24 May 2004, and shown in false color, shows the evolution of the particular chromospheric loop first identified by SUMER.The white vertical line in the center frame shows the position of the SUMER slit during this observation. North is upward and west is to the right in this figure, as well as in Figs. 2, 3, and 10.}
\label{loop}
   \end{figure}

Figure \ref{loop} shows the overall evolution of the feature in 1600 \AA\ images. A loop emanates from a pre-existing bright patch-like area, and has brightenings at both footpoints early on (visible, \emph{e.g.}, in the 09:17:03 UTC panel).  In the frame at 09:17:44 UTC, the west end of the loop dominates the intensity, and in the following frame the entire loop is strongly illuminated.  By 09:20:03 UTC, most of the loop has faded, but the west-side footpoint still remains bright.

 \begin{figure}   

\centering
\includegraphics[width=1.0\textwidth]{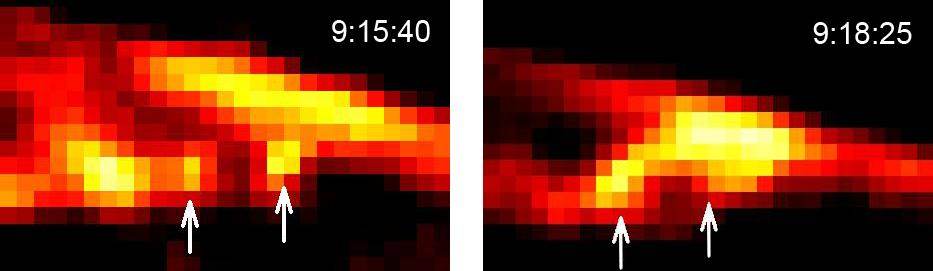}
\caption{Two TRACE images taken in the continuum wavelength 1600 \AA\ on 24 May 2004, and shown in false color, designates the position of the chromospheric loops' two footpoints. The diagonal streak is a feature related to the limb.}
\label{footpoints}
   \end{figure}

Overall, in 1600 \AA\ images the lifetime of the loop is approximately 300 s.  By counting the pixels between the footpoints (as seen in Figure \ref{footpoints}), the distance is calculated to be approximately 2000 km. Although the tilt of the loop cannot be determined exactly, estimates place the length of the loop at approximately 3000 km. Based on these measures and intensity variations, we can estimate velocities for flows along the loop to be approximately $\pm$10 km s$^{-1}$. However, the material appears to travel from the right footpoint to the left footpoint and back to the right footpoint. In this case, the speed of the material would be approximately $\pm$20 km s$^{-1}$.

\begin{figure}[!h]   
\centering
\includegraphics[width=1.0\textwidth]{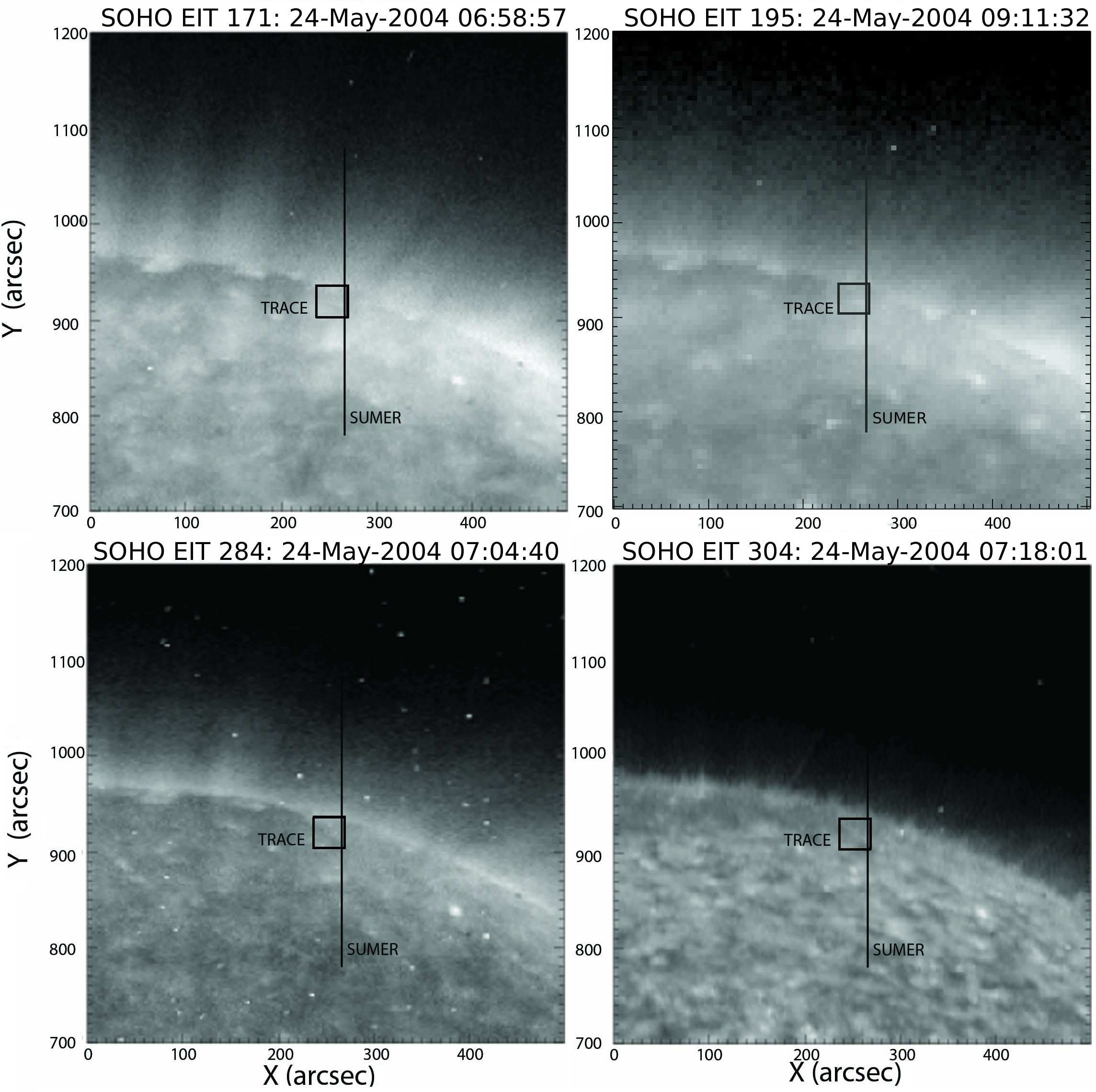}
\caption{The limb region we observed in, respectively 171 \AA\ (Fe {\sc xi} and Fe {\sc xi}, $T$$\approx$1 MK), 195 \AA\ (Fe {\sc xii}, $T$$\approx$2 MK), 284 \AA\ (Fe {\sc xv}, $T$$\approx$2.5 MK), and 304 \AA\ (He {\sc ii}, $T$$\approx$60,000 K, blended with weaker Si {\sc xii}). No sign of a coronal jet was seen on the sequence of EIT images.}
\label{eit}
   \end{figure}

TRACE did not observe this region in any of its coronal filters (195 \AA, 284 \AA, or 171 \AA), and so we cannot use it to determine the coronal environment of the feature.  There are, however, coronal data from SOHO’s Extreme-ultraviolet Imaging Telescope (EIT) available, in 195 \AA\ from around the time of the transient loop, at 08:35 UTC, 08:46 UTC, 09:12, and 09:23 UTC, and in 171 \AA, 195 \AA, and 284 \AA\ between 06:59 UTC and 07:12 UTC, all on 24 May 2004.  Due to the relatively poor time cadence, EIT does not have an image showing the transient loop itself.  We can nonetheless use the EIT images to identify the approximate location of the loop in the corona.  It occurred very near the western boundary of the northern coronal hole as shown in Figure \ref{eit}.  Since we do not see the event in the EIT images, and because our view of the region around the coronal hole boundary is partially impeded by foreground bright coronal material, it is not possible to say whether the event occurred precisely at the boundary, or just inside or outside of it.

Though recent observations with STEREO/EUVI (Patsourakos \emph{et al.}, 2008), and with EUVI in combination with \emph{Hinode}/XRT (Moore \emph{et al.}, 2010), show some polar coronal jets in which cool plasma as well as hot coronal plasma erupts, our inspection of the EIT images did not reveal any such eruptions.

\subsection{SUMER Data} 
  \label{S-SUMER}

\begin{figure}[!h]  

\centering
\includegraphics[scale=.4, angle=90]{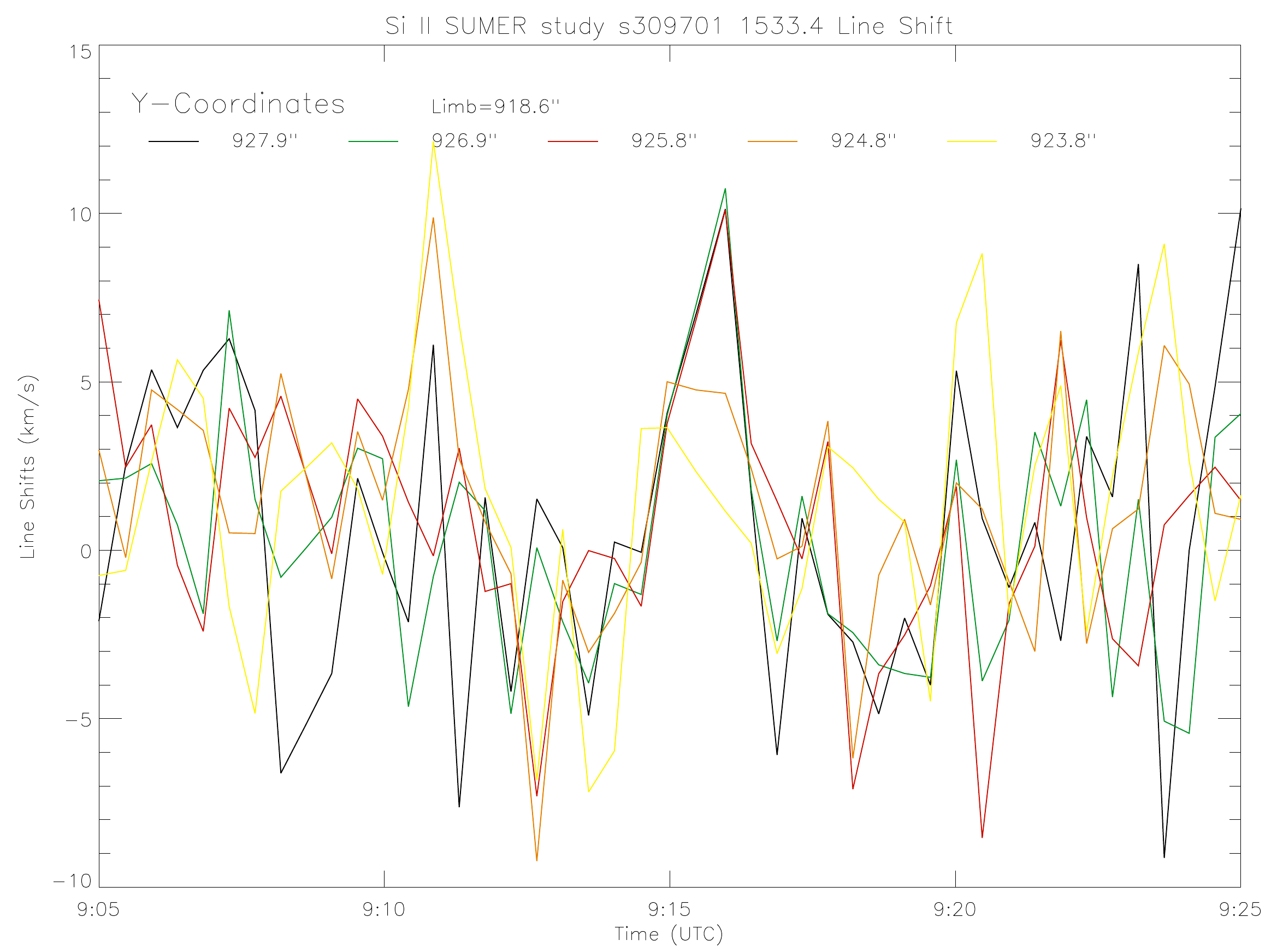}
\caption{Line-shift graph ($\pm\approx$10 km s$^{-1}$ from average, measured from the center of gravity of the line profile) exhibiting the strong coherent redshifting of Si {\sc ii} emission at approximately 9:15 UTC.}
\label{SII}
   \end{figure}

\begin{figure}[!h]   

\centering
\includegraphics[scale=.4,angle=90]{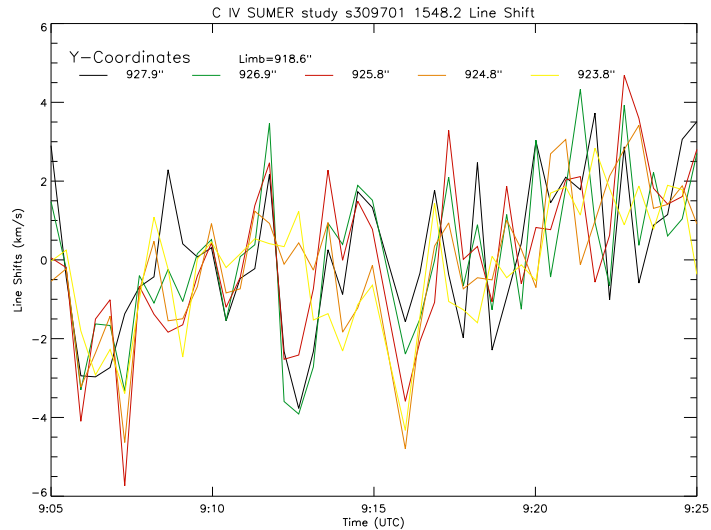}
\caption{Line-shift graph ($\pm\approx$5 km s$^{-1}$ from average, measured from the center of gravity of the line profile) exhibiting the strong coherent blueshifting of C {\sc iv} emission at approximately 9:15 UTC.}
\label{CIV}
   \end{figure}

Two or three minutes before the loop is observed in TRACE, strong redshifts of approximately 10 km s$^{-1}$ were detected in the SUMER data in Si {\sc ii} (Figure \ref{SII}) along with simultaneously strong blueshifts of approximately $-$5 km s$^{-1}$ in C {\sc iv} (Figure \ref{CIV}). Concurrent radiance measurements (from integration after continuum subtraction, as described in Section 2) taken with SUMER show a significant increase in Si {\sc ii} radiance (Figure \ref{radiance}) along with a significant decrease in C {\sc iv} radiance (Figures \ref{radiance} and \ref{arrow}).

Considering these findings in terms of the TRACE loop observations, the increase in
Si {\sc ii} radiance could be explained by relatively cool material traveling along magnetic
field lines higher in the solar atmosphere, effectively increasing its radiance at those
y-coordinates. But the decrease in C {\sc iv} radiance could be understood by a cooling
of hotter material, decreasing the C {\sc iv} radiance and possibly moving into the Si {\sc ii}
passband, increasing its radiance. Though Si {\sc ii} may be optically thicker than C {\sc iv}, we
do not expect Si {\sc ii}-emitting material to mask the C {\sc iv}-emitting material, even if the
Si {\sc ii} material has been carried higher in the atmosphere.

\begin{figure}  
\centering
\includegraphics[scale=.7]{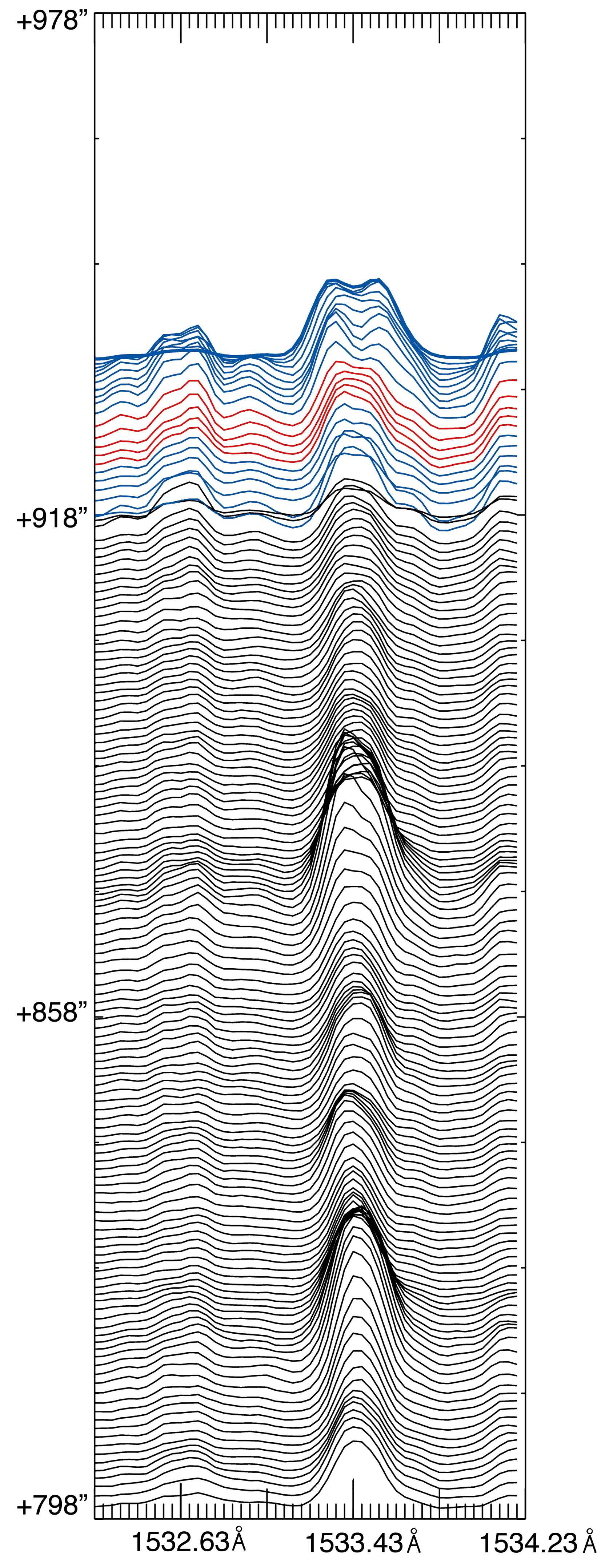}
\caption{A series of 138 spectra at different heights from the limb downward (south from the north polar region) at 1 arcsec per profile, onto the disk, with the top 20 profiles above the limb colored blue. The five red profiles are the spectra that correspond to the line shift plot in Figure \ref{SII}. Each profile is an average over all observations taken during the sit-and-stare phase, during which the
position of the slit does not change. The low pair of scans at $y$$\approx$+918'' results from a detector fault.}
\label{spectra}
   \end{figure}

\begin{figure}   

\centering
\includegraphics[width=1.0\textwidth]{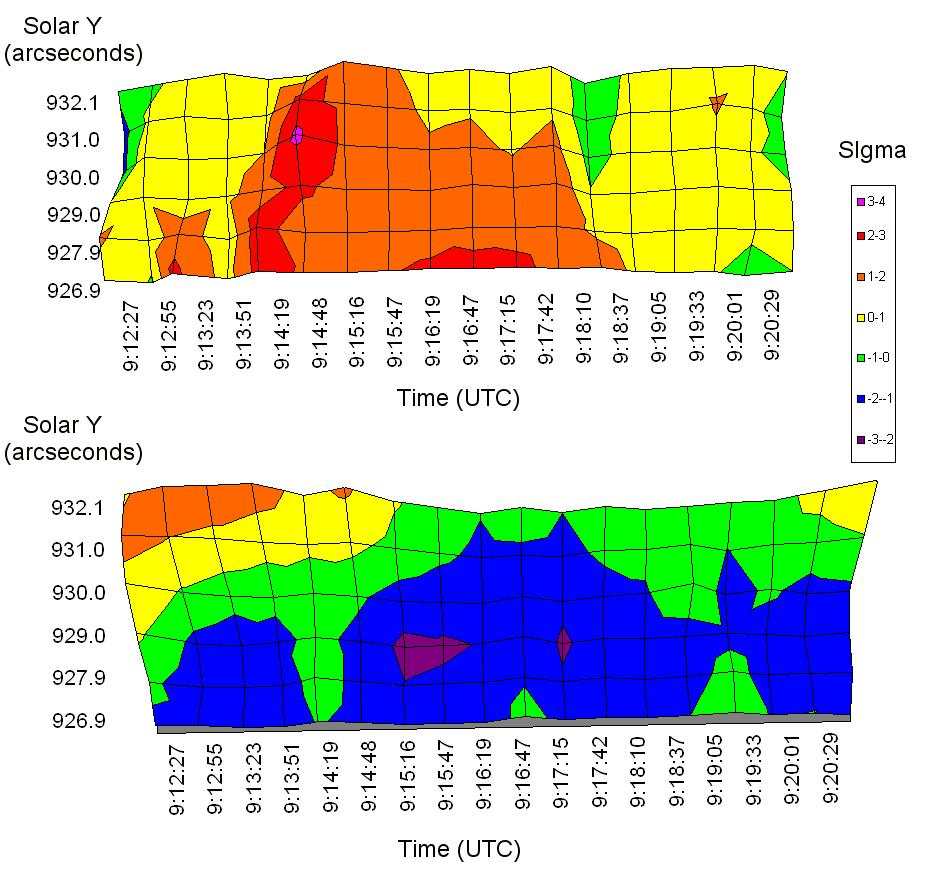}
\caption{The different colors show the value of radiance-difference (measured radiance from average radiance) in units of  $\sigma$ for a given height in arcseconds and time, where $\sigma$
is the calculated standard deviation from the mean for the radiance of
a certain wavelength at a specific $y$-coordinate. Si {\sc ii} is represented in the top graph
while C {\sc iv} radiance is below. The observations show a simultaneous increase in Si {\sc ii} radiance and a coherent Si {\sc ii} velocity redshift. Oppositely, the observations show
a simultaneous decrease in C {\sc iv} radiance and a coherent C {\sc iv} velocity blueshift.
Our maximum velocity in Figures \ref{SII} and \ref{CIV} differs by approximately four pixels from the
maximum intensity-difference shown in this figure, perhaps a result of our slit, which
was arbitrarily placed, not being exactly on the position of maximum velocity.}
\label{radiance}
   \end{figure}

\begin{figure}   

\centering
\includegraphics[width=1.0\textwidth]{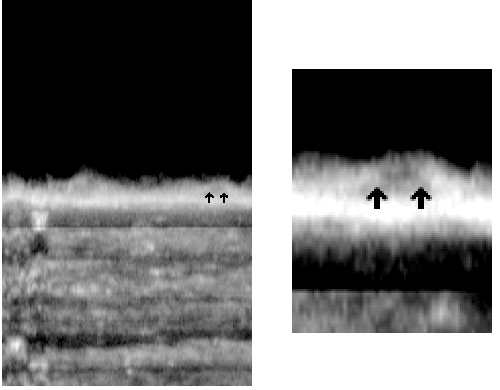}
\caption{The SUMER scan in C {\sc iv} (1548 \AA). The left image shows the entire 90 min $xt$-scan while the right image shows a closer view of the scan where the chromospheric loop occurred. The arrows in each image are pointing to the beginning and ending of the loop observed with TRACE, from 9:15 to 9:20 UTC at $y$=929 arcsec. Notice the darkening of the C {\sc iv} material between the two arrows. This occurrence is reflected in the bottom graph of Figure \ref{radiance}.}
\label{arrow}
   \end{figure}

The spectral line shape for Si {\sc ii}, which is centered at 1533.43 \AA\, is non-Gaussian
outside the limb (Figure \ref{spectra}), which has alternative explanations that follow a long-running controversy. (a) Such structure could be a radiative-transfer effect with
self-absorption, as one of us (IED) has found for other SUMER Si {\sc ii} center-to-limb
observations, showing that Si {\sc ii} is not optically thin. (b) The splitting of the peaks
in the spectrum could be an effect of a superposition of spicules with velocities
showing for two different emitting regions (Pasachoff, 1970; Pasachoff and Zirin,
1971), a physical splitting that is endorsed in part by observations that spicules
frequently exhibit lateral motions. The dip at the limb cannot merely be a blend
with Si {\sc i} 1533.41 \AA\ (Curdt, 2008), since Si {\sc i} vanishes at and below the limb.  Decomposition shows, at 9:16:19 UTC, for example, a stronger profile at 1533.6 \AA\ (about 0.07 \AA\ redward of the Gaussian peak that had been fit to the non-Gaussian line profile) and a profile weaker approximately by a factor of 3 peaking at 1533.3 \AA. The velocity of the brighter Si {\sc ii} component is approximately 20 km s$^{-1}$ while the velocity of the weaker component is approximately $-$25 km s$^{-1}$. The Gaussian peak that was fit to the non-Gaussian line profile gives a velocity of 10.2 km s$^{-1}$. C {\sc iv} has a single, Gaussian profile, and is measured at approximately $-$4.5 km s$^{-1}$. The Ne {\sc iii} line at 770 \AA\, observed in second order, shows essentially no velocity shift (approximately 0.9 km s$^{-1}$). 

Figure \ref{radiance} illustrates the differences between values of the radiance at a given $y$-coordinate with the average value of the radiance at that $y$-coordinate. The greater the magnitude of the multiple of $\sigma$, the greater the probability that the measured increase or decrease in radiance is a significant change rather than a normal statistical variation. Therefore, the most significant increase in Si {\sc ii} radiance, 3.21 (as shown in the top of Figure \ref{radiance}), occurs at $y$=930 arcsec and at time 9:14:48 UTC. In contrast, the most significant decrease in C {\sc iv} radiance, $-$2.26 (as shown in the bottom of Figure \ref{radiance}), occurs at $y$=929 arcsec and at time 9:15:16-9:15:47 UTC. This can also be seen in Figure \ref{arrow} where the C {\sc iv} radiance decreases during the lifetime of the chromospheric loop. These times correspond within about one minute to the times of the strongest lineshifts given in Figures \ref{SII} and \ref{CIV}. Moreover, the times for both the peak lineshifts and the strongest SUMER irradiances are earlier than the start of the main phase of the loop flows appearing in the TRACE 1600 \AA\ images.

\begin{figure}   

\centering
\includegraphics[width=1.0\textwidth]{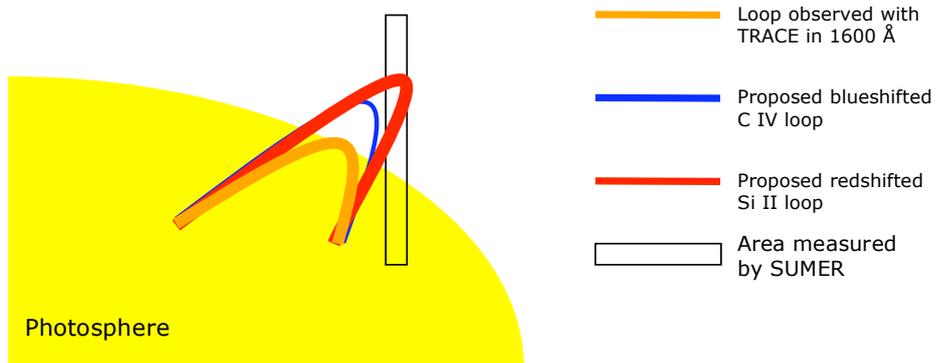}
\caption{Illustration of multiple hypothetical loops measured with SUMER alongside chromospheric loop observed with TRACE.}
\label{diagram}
   \end{figure}

\section{Discussion}

Figure \ref{diagram} shows a schematic representation of our view of the relation between
the solar features and the observations. Because the SUMER slit was placed
slightly to the right of the chromospheric loop, which was visible with TRACE,
our spectroscopic observations did not directly measure the visible chromospheric
loop but instead the associated material traveling higher in the solar atmosphere.
Because of the timing of the blue- and redshifted material measured with SUMER
and the chromopheric loop observed with TRACE, we believe that the two events
are related and are just different observations of the same event. Because the
proposed Si {\sc ii} loop was measured with a red-shifted velocity, the material must
have originated from the footpoint in the foreground and travelled to the footpoint in the
background. Oppositely, the C {\sc iv} loop was measured with a blue-shifted velocity
and must have travelled from the background footpoint to the foreground footpoint.

The velocity measured in C {\sc iv} as a blueshift is lower in absolute value than the
velocity of the Si {\sc ii} feature measured as a redshift, which may be a result of a hotter,
C {\sc iv}-dominated strand being measured less directly than a cooler, Si {\sc ii}-dominated
strand; that is, if the cooler, Si {\sc ii}-dominated strand is more centered in the SUMER
slit, its velocity would be stronger than if only the edge of the hotter C {\sc iv}-dominated
strand were measured with SUMER. Further, we measure the greatest difference
in Si {\sc ii} radiance, in units of $\sigma$, at $y$=930 arcsec while the greatest difference in C
{\sc iv} radiance is at $y$=929 arcsec, providing evidence that there were material flows
at different temperatures for slightly different altitudes. The blue- and red-shifted
velocities show that the C {\sc iv} and Si {\sc ii} might be moving in opposite directions.

We probably were not observing the boundary of the feature. Although it continues
to decrease in brightness the higher in $y$-coordinate we go (as to be expected) at
9:16:19 UTC, we are still seeing strong shifts throughout a range of $y$-coordinates. At
y=930 arcsec at 9:16:19 UTC, there is still a 4 km s$^{-1}$ redshift. Although the stronger
redshifts are below it, this increase is still strong and may be a result of the material at
lower $y$-coordinates being accelerated before the material in the higher coordinates is.
Or, perhaps the material in the lower coordinates is being accelerated into the higher
coordinates, accounting for the delay.

The separation of the maximum increase in Si {\sc ii} radiance and the strongest Si {\sc ii}
redshift is approximately 4 arcsec and 91 s. But the increasing velocities of the
material in the higher coordinates, albeit not as fast as those in the lower coordinates,
and the delay of the accelerated material in the lower coordinates being moved
towards the higher coordinates, might serve to explain this discrepancy.

Also, even though the peak velocity shifts occur at 9:16:19, Figures 4 and 5 show that
the material begins to coherently accelerate at approximately 9:14:48, the time when
the greatest increase in radiance is occurring. Also, if one looks further back, the
material appears to start positively accelerating at approximately 9:13:51, although it
was still being blueshifted.

We conclude that there was a relatively quick event that caused the increased Si {\sc ii}
radiance. Although the pink circle in the radiance graph (Figure 7) shows the greatest increase in radiance, there are still significant increases in radiance at the lower $y$-
coordinates as well. Starting from $y$=926.9 arcsec, one can see a significant increase
in Si {\sc ii} radiance at approximately 9:14:19 UTC, about 30 s before the strongest
increase in radiance. Shortly thereafter, the regions in the lower $y$-coordinates start to
be redshifted and reach their peak velocity at 9:16:19. The material at this point is still
brighter than before the event, although the intensity has tapered off.

We further conclude that this increase in brightness, starting at lower coordinates
and working up to higher coordinates, and the redshifts, which also start at lower
coordinates and continue into the higher coordinates, seemingly following the
radiance increases, are indeed correlated. The C {\sc iv} radiance decrease also occurs a
few arcseconds above the greatest C {\sc iv} blueshifting measurements.

The close proximity and timing of the SUMER feature and the TRACE transient
loop give us confidence that both instruments are observing the same feature; the
SUMER slit happened to measure the area slightly to the right of the chromospheric
feature. In addition, the concurrent and close proximity of the blue- and redshifting
of atmospheric material indicates that the two flows are most likely connected and
have the same origin. Also, because the time between the beginning and end of
the velocity shifts are much shorter than the time that the TRACE chromospheric
feature is visible, it is possible that the velocities of the C {\sc iv} and Si {\sc ii} changed
abruptly just before the feature became visible in the chromosphere. One explanation
for the red- and blueshifted measurements is that the C {\sc iv} and Si {\sc ii} flowed along
magnetic field lines above the TRACE loop. The observations show that the material
velocities start to coherently shift in velocity before the TRACE loop is visible.

It is not clear what other solar features correspond to our observed transient loop
event. As this event occurs very near, and perhaps inside of the polar coronal hole,
it could be a counterpart to the base of an X-ray jet (\emph{e.g.}, Strong \emph{et al.}, 1992; Shimojo
\emph{et al.}, 1996; Cirtain \emph{et al.}, 2007; Culhane \emph{et al.}, 2007; Kim \emph{et al.}, 2007). The X-ray jets
that occur in polar regions have a strong tendency to occur inside of coronal holes
(Savcheva \emph{et al.}, 2007), which is, as discussed above, approximately where our feature
occurred. They have lifetimes clustered around a few hundred seconds (Savcheva \emph{et al.}, 2007), and that is comparable to the lifetime of our feature. The erupting polar
plume reported by Pasachoff \emph{et al.} (2008) at an eclipse may be similar. The jets are
suspected to be due to magnetic reconnection between an emerging magnetic bipole
and a pre-existing open magnetic field line, and such a scenario has been modeled by
Shibata \emph{et al.} (1992).

Our loop and flows could correspond to this jet picture, with the SUMER flows
corresponding either to jet material or to newly-reconnected hot magnetic loops at
the jet's base; both features would be expected to be invisible in the cool TRACE
1600 \AA\ images. As the hotter new loops cooled, they would eventually appear in the
TRACE 1600 \AA\ passband, and likely as our loop feature. The delay in time between
the SUMER flows and radiance peaks, both of which are detected as relatively hot
emission, and the loop in TRACE would be consistent with this, assuming that TRACE is observing only lower chromospheric emissions (the 1600 \AA\ channel can detect Lyman-$\alpha$ emission in hot flares, but it is unclear whether enough such
emission to dominate the cooler emissions would have been produced in the feature
we observe). Sterling and Moore (2005) saw a mound-like structure in TRACE 1600 \AA\ images that was similar to the transient loop feature here. They attributed that feature
to reconnection during the preflare phases of a solar eruption. Similarly, reconnection
may produce the transient loop seen in 1600 \AA\ here.

The observations presented here took place after the end of the \emph{Yohkoh} solar mission,
and prior to the \emph{Hinode} mission, and so soft X-ray data from those respective
satellites were not available. Data from the Solar X-ray Imager (SXI) aboard
GOES 12 (Figure 10) does not show any remarkable activity along the western
border of the northern coronal hole during the time span of the transient event.

\begin{figure}   

\centering
\includegraphics[scale=.4]{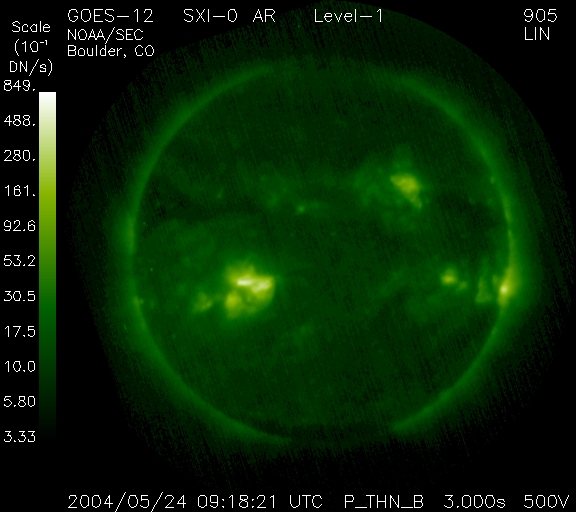}
\caption{Data retrieved from SXI
aboard the GOES 12 spacecraft
does not show any apparent activity
along the western border of the
northern coronal hole where the
chromospheric loop was observed.
The large FOV and low resolution
of the SXI would make observation
of a small chromospheric loop
extremely difficult.}
\label{SXI}
   \end{figure}

In future work, we would like to compare observations with a fast-cadence chromospheric imaging observatory, such as SDO/AIA observing with cooler filters, and a similar fast-cadence X-ray telescope, like \emph{Hinode}'s XRT, simultaneously on the solar limb. So far, most such joint studies have been done with TRACE observing with filters sensitive to hotter
temperatures. With such an observing program, which is automatically available with Solar Dynamics Observatory, we would be able to establish whether these small chromospheric loops, though at a much cooler temperature, are connected to the X-ray jet phenomenon.

\begin{acks}
We thank William A. Jacobson for his collaboration with data reduction at Williams
College. David Butts, Joseph Gangestad, Owen Westbrook, Jennifer Yee, Megan Bruck, and Anne Jaskot were students who participated in the observations on site during our three years of observing with the Swedish 1-m Solar Telescope, and Kamen Kozarev participated in earlier data reduction. We thank Leon Golub and Edward DeLuca (Harvard-Smithsonian Center
for Astrophysics), and Jonathan Cirtain (now at NASA's Marshall Space Flight
Center) for consultations on the TRACE data reduction. We thank Steven P. Souza of
Williams College's Astronomy Department for computing assistance and advice. We thank Klaus Wilhelm for advice and for comments on the manuscript.

Our work was funded in part by NASA grants NNG04GF99G and NNG04GK44G from the Solar Terrestrial Program and grants NNM07AA01G and NNX10AK47A from NASA's Marshall Space Flight Center. Tingle's participation was sponsored by a
grant from the Research Experiences for Undergraduate Program of the National
Science Foundation to the Keck Northeast Astronomy Consortium, formerly
sponsored by the Keck Foundation. A.C.S. was supported by funding from NASA's
Office of Space Science through the Living with a Star, the Solar Physics Supporting
Research and Technology, and the Sun-Earth Connection Guest Investigator Programs.
SOHO is a joint project of the European Space Agency and NASA. TRACE is a project of NASA, with its telescope from the Smithsonian Astrophysical
Observatory and overall direction from Lockheed Martin Solar and Astrophysics
Laboratory.
\end{acks}




\end{article} 

\end{document}